\documentclass[12pt,preprint]{aastex}
\begin{document}
\title{A catalog of planetary nebulae in
the elliptical galaxy NGC 4697\altaffilmark{1}}

\author{R. H. M\'endez, A. M. Teodorescu, and R.-P. Kudritzki}
\affil{Institute for Astronomy, University of Hawaii, 
                 2680 Woodlawn Drive, Honolulu, HI 96822}
\email{mendez@ifa.hawaii.edu}

\altaffiltext{1}{The data presented herein were obtained at the European 
Southern observatory, Cerro Paranal, Chile, Program ESO 63.I-0008.}

\begin{abstract}

We present a catalog of 535 planetary nebulae discovered in the 
flattened elliptical galaxy NGC 4697, using the
FORS1 Cassegrain spectrograph of the Very Large Telescope of the
European Southern Observatory at Cerro Paranal, Chile.
The catalog provides positions ($x, y$ coordinates relative
to the center of light of NGC 4697, as well as $\alpha, \delta$), 
and, for almost all PNs, the magnitude $m$(5007) and the heliocentric 
radial velocity in km s$^{-1}$.

\end{abstract}


\section{Introduction}

We present a catalog of 535 planetary nebulae (PNs) 
discovered in the flattened elliptical galaxy NGC 4697, using the
FORS1 Cassegrain spectrograph of the Very Large Telescope of the
European Southern Observatory at Cerro Paranal, Chile.
The observations and scientific results have been described in 
M\'endez et al. (2001, in what follows Paper I). 
The interpretation of the most important result, which can be
described as a Keplerian decline of line-of-sight velocity 
dispersion as a function of angular distance from the center
of the galaxy, has been quite difficult. Several theoretical groups
are working on the development of better numerical models, including 
three-integral models better adapted to a flattened dynamical 
system like NGC 4697. We have decided to publish the catalog now, 
so as to make 
it available to all interested groups. We have recalculated the 
radial velocities in one of the observed fields, using a different 
procedure than the one described in Paper I. After confirming that
no important changes were necessary, we have combined the two sets 
of measurements into one final catalog. The catalog provides 
positions ($x, y$ coordinates relative
to the center of light of NGC 4697, as well as $\alpha, \delta$), 
and, for almost all PNs, the magnitude $m$(5007) and the heliocentric 
radial velocity in km s$^{-1}$.

We also provide a list of peculiar, mostly emission-line objects that 
are not planetary 
nebulae, for eventual future studies. In section 2 we describe the 
radial velocity recalculations mentioned above, and compare the new 
versus the old velocities. Section 3 contains the catalog, and Section 
4 is devoted to the peculiar objects we detected.

\section{Radial velocity recalculations}

Planetary nebulae in NGC 4697 were discovered using a combination of 
onband images, offband images, and onband $+$ grism dispersed
images. For details, please refer to Paper I (M\'endez et al. 2001).

In Paper I, the registration of grism images was performed using the
dispersed images of stars. This introduces a problem, already 
described in Paper I, Section 3: since the on-band filter 
transmission curve shifts in wavelength as a function of ambient 
temperature, it turns out that the positions of the spectral 
segments of continuum sources in the grism images are dependent on 
temperature, while the positions of emission-line sources are not 
affected. One of the two fields observed in Paper I, namely field W,
was obtained by combining images taken in two different years at 
somewhat different temperatures; since the temperature-dependent 
positions of dispersed stars were used for the registration, the PN
positions in different years were slightly different. 

Although this problem could be detected and corrected, using the
positions of bright PNs in the individual reference image for field 
W, we always wanted to redo the registration for this field using 
directly the brightest PNs visible in all the 
individual images, instead of using the dispersed star segments, 
because, using the PNs, the temperature effect is completely avoided.
We have finally done so. For the new registration we used 30 bright
PNs visible in the individual W onband exposures.
Once the new combined W dispersed image was 
produced, we remeasured the positions of all PNs previously detected
in the W field, and recalculated all their radial velocities.
Fig. 1 shows the comparison of old versus new velocities for 
all PNs in the W field. Fig. 2 shows the difference between
new and old W velocities as a function of the new W velocities.
The agreement is satisfactory, but
there is a small systematic difference of about 20 km s$^{-1}$.
This offset is also visible in Figs. 3 and 4, where the differences 
are plotted respectively as a function of the $m$(5007) magnitudes, 
and of the E velocities (where available). Note in Fig. 3 how the
differences get larger for fainter PNs, as expected, 
since the quality of the radial velocities depends on the quality 
of the position measurements.

The systematic difference between old and new W velocities is unexpected, 
because the new measurements are not supposed to be less reliable than the 
old ones. We have not found an explanation for the offset. We could have 
ignored the new measurements and kept the old ones, but we feel it is 
more honest to consider old and new W measurements as independent 
(because a new combined image was generated and measured) 
and equally valid. Consequently, and since the offset is small in 
comparison to errors of about 40 km s$^{-1}$, as estimated in Paper I,
we decided to average the two radial velocities for each 
PN in the W field. The resulting W radial velocities were then averaged
with the corresponding velocities from detections in the E field, when
available. 

Fig. 5 shows the velocities of 531 PNs as a function of 
their $x$-coordinates relative to the center of NGC 4697; this is
essentially the same as Figure 20 in Paper I, confirming that there 
is no significant change in the radial velocity distribution.
In other words, all discussions and conclusions in Paper I remain
valid.

\section{The PN catalog}

In Table 1 we list the following: identification numbers for both E 
and W fields; ($x, y$) coordinates, in pixels, relative to the optical 
center of NGC 4697; J2000 equatorial coordinates; Jacoby magnitudes
$m$(5007); and heliocentric radial velocity in km s$^{-1}$. 

It may be useful to point out that the identification numbers do not 
follow a continuous sequence; they were assigned in arbitrary order 
as the PNs were being discovered, and the numbers for objects in 
field W start at 1001 in order to avoid any confusion with E field 
numbers, which are all below 1000. A few numbers are missing because
the corresponding objects were rejected as PNs. There is also a 
discontinuity between W1217 and W1501. We prefer to keep these original 
identification numbers unchanged, for our own reference in future 
studies and because some of these numbers have already been used in 
another publication, namely the chemical abundance study (M\'endez 
et al. 2005). An identification number $-1$ was assigned in Table 1 
to objects not present in the corresponding field.

The original on-band and off-band images (fields E and W)
were oriented with the $x$-axis in the direction of the major axis. 
The combined images are available in fits format upon request. 
We have not listed the 
PN pixel coordinates in each combined on-band image, because they can 
be easily obtained from the ($x, y$) values in the catalog, knowing 
that the ($x, y$) pixel coordinates of the center of NGC 4697 in the 
E and W images are, respectively, (1543, 979) and (373, 968).

The J2000 equatorial coordinates were calculated using a set of 
astrometric programs written by David Tholen and kindly provided 
by Fabrizio Bernardi. A first program, when given an input file 
with field and camera parameters, identifies all the USNO-B1
catalog stars available within the desired field and produces a list
of reference stars with rough estimates of their position in the chip.
The next programs make an improved centroid fitting for reference 
stars, producing an output file with a list of reference stars 
with pixel coordinates $x, y$ \/ accurate to a few hundredths
of a pixel. The last program performs the final astrometric fit,
rejecting outliers and iterating, and provides the PN equatorial 
coordinates with uncertainties of about 0.2 arcsecond. We list 
those final equatorial coordinates in Table 1. The uncertainty in 
RA and declination close to the center of NGC 4697 is probably a bit
larger, about 0.5 arcsecond, because there are almost no reference 
stars near that position.

The definition of Jacoby magnitudes $m$(5007) can be found in Paper I,
Section 4. Or see Jacoby (1989).
Typical uncertainties for these magnitudes are 0.1 and 0.2 
mag for PNs brighter and fainter than $m$(5007) = 26.5, respectively. 

Finally, in Paper I we estimated that the heliocentric radial 
velocities have uncertainties of about 35 or 40 km s$^{-1}$.
For both magnitudes and heliocentric radial velocities, 
a value of $-1$ in Table 1 indicates that the corresponding 
quantity could not be measured.

\section{A list of other sources in the field of NGC 4697}

The PN search method, based on taking onband, offband and 
(onband$+$grism) images, is affected by some contamination 
from other sources. Since some of these might become 
interesting for other purposes, we have decided to list them. 
These objects can be assigned to one or another of the 
following groups:

(1) Emission-line regions within an extended source, presumably 
star-forming galaxies at 
such redshifts that some emission line falls into the on-band 
filter transmission curve (e.g. [O II] $\lambda$3727 at $z=0.35$).

(2) Point sources visible not only in the onband and grism images,
but also in the offband image. These could be high-redshift quasars or 
star-forming regions (e.g. showing Lyman $\alpha$ redshifted into the
onband filter at $z=3.1$), 
or perhaps even a PN within a globular cluster that belongs to 
NGC 4697 (one such globular cluster was reported in NGC 5128
by Minniti and Rejkuba 2002). The object is rejected as a PN because 
it is visible in the offband image; but of course it would have to be 
reaccepted as PN if the last interpretation were to be confirmed.

(3) Point sources visible in the onband and grism images, and 
invisible in the offband image, but having a radial 
velocity incompatible with NGC 4697 if the emission line is assumed 
to be [O III] $\lambda$5007. Three such sources were reported 
in Paper I, and we include them here.

(4) Sources stronger in the offband image, presumably star-forming 
galaxies or quasars with some emission line redshifted into the 
offband filter.

(5) A few sources were found to be stronger in the onband image than 
in the offband image, but no emission line was detectable in the grism 
image. Perhaps these are variable stars.

Table 2 lists a total of 20 sources, providing the $x, y$ pixel 
coordinates relative to the center of NGC 4697,
the J2000 equatorial coordinates, and a brief description.
A search within the Simbad reference database has not produced any
coincidence of objects in Table 2 with known sources. Object 10 in
Table 2 is very close (a few arcseconds) to object 29 in the list of
Chandra X-ray sources of Sarazin, Irwin, \& Bregman (2001), but given
the small coordinate uncertainties we believe these two sources are
not related.
 
\acknowledgements
The work presented here has been supported by the National Science 
Foundation under Grant No. 0307489. We would like to thank Fabrizio
Bernardi for his help using David Tholen's astrometric programs 
for the calculation of J2000 equatorial coordinates.

\clearpage

\begin{deluxetable}{rrrrrrrrrrrr}
\tablecaption{Planetary nebulae in NGC 4697 \tablenotemark{a} \label{tbl-1}}
\tablewidth{0pt}
\tabletypesize{\tiny}
\tablehead{
\colhead{Id, E} & \colhead{Id, W} & \colhead{$x$, G} & \colhead{$y$, G} & 
\colhead{\ } & \colhead{$\alpha$} & \colhead{(2000)} & 
\colhead{\ } & \colhead{$\delta$} & \colhead{(2000)} &
\colhead{$m$(5007)} & \colhead{Helioc. RV}
} 
\startdata

      7 &   1001 &   -295 &   -801 &    12 &   48 &   44.02 &    -5 &   50 &    2.4 &    26.6 &    1228 \\
      9 &   1070 &     41 &   -894 &    12 &   48 &   40.47 &    -5 &   50 &   47.8 &    26.9 &     980 \\
     20 &   1002 &   -254 &   -670 &    12 &   48 &   42.78 &    -5 &   49 &   42.0 &    27.3 &    1354 \\
     21 &   1004 &   -209 &   -581 &    12 &   48 &   41.73 &    -5 &   49 &   30.0 &    26.3 &    1283 \\
     22 &   1003 &   -181 &   -713 &    12 &   48 &   42.14 &    -5 &   49 &   56.1 &    27.4 &    1283 \\
     23 &   1071 &   -110 &   -707 &    12 &   48 &   41.25 &    -5 &   50 &    1.1 &    27.4 &    1140 \\
     24 &   1072 &    -90 &   -667 &    12 &   48 &   40.78 &    -5 &   49 &   55.5 &    26.5 &    1266 \\
     25 &   1073 &     38 &   -596 &    12 &   48 &   38.82 &    -5 &   49 &   53.6 &    26.1 &     995 \\
\enddata 
\tablenotetext{a}{The $x$, G and $y$, G pixel coordinates have their origin 
at the center of light of NGC 4697. The $x$, G coordinate is defined along 
the major axis of NGC 4697. The units of Right Ascension are hours, minutes
and seconds; the units of Declination are degrees, arcminutes, and arcseconds.
Heliocentric radial velocities are given 
in km s$^{-1}$. In columns 1 \& 2 a value of $-1$ indicates that the 
object is not present in the 
corresponding field. In columns 7 \& 8 a value of $-1$ indicates that the 
corresponding quantity could not be measured.}
\end{deluxetable}

\begin{deluxetable}{rcrrrrrrrrr}
\tablecaption{Some additional objects in the field of NGC 
                                   4697 \tablenotemark{a} \label{tbl-2}}
\tablewidth{0pt}
\tabletypesize{\tiny}
\tablehead{
\colhead{Object} & \colhead{Field} & \colhead{$x$, G} & \colhead{$y$, G} & 
\colhead{\ } & \colhead{$\alpha$} & \colhead{(2000)} & 
\colhead{\ } & \colhead{$\delta$} & \colhead{(2000)} &
\colhead{Notes}
} 
\startdata

      1  & E,W &   -136 &   -278  & 12  &  48  &  39.14  &  -5  &  48 &  41.4 &  b \\
      2  & E,W &    -93 &    239  & 12  &  48  &  35.68  &  -5  &  47 &  11.3 &  c \\
      3  & E,W &    190 &   -449  & 12  &  48  &  36.15  &  -5  &  49 &  40.0 &  b \\
      4  & E,W &     32 &   -270  & 12  &  48  &  37.05  &  -5  &  48 &  54.2 &  c \\
      5  & E,W &     96 &    805  & 12  &  48  &  30.14  &  -5  &  45 &  45.0 &  b \\
      6  & E,W &    405 &   -684  & 12  &  48  &  34.86  &  -5  &  50 &  40.6 &  c \\
      7  & E,W &    297 &    193  & 12  &  48  &  31.20  &  -5  &  47 &  52.8 &  c \\
      8  & E,W &    359 &    115  & 12  &  48  &  30.89  &  -5  &  48 &  12.2 &  d \\
\enddata 
\tablenotetext{a}{The $x$, G and $y$, G coordinates have their origin at 
the center of light of NGC 4697. The $x$, G coordinate is defined along 
the major axis of NGC 4697. The units of Right Ascension are hours, 
minutes and seconds; the units of Declination are degrees, arcminutes, 
and arcseconds.}
\tablenotetext{b}{Emission-line region within an extended source (galaxy).}
\tablenotetext{c}{Point source in onband and grism+onband, 
                                visible but not stronger in offband.}
\tablenotetext{d}{Point source in onband, stronger in offband, 
                                           absent in grism+onband.}
\tablenotetext{e}{Point source in onband, visible but weaker in offband, 
                                 absent in grism+onband. Variable star?}
\tablenotetext{f}{Point source in onband and grism+onband, not visible 
    in offband, but wrong velocity for NGC 4697 if identified as 
    [O III] $\lambda$5007.}

\end{deluxetable}

\clearpage

\figcaption[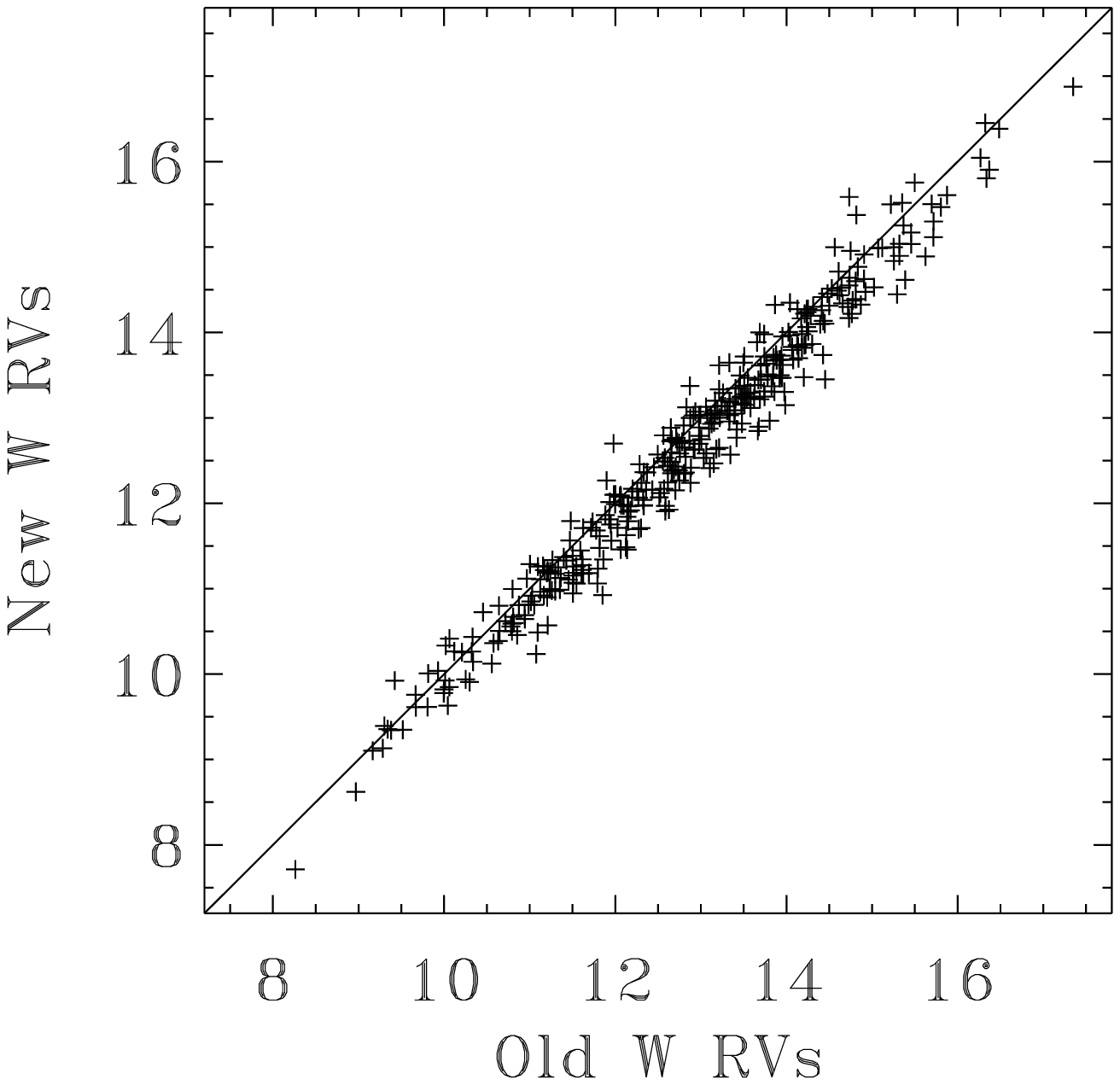]{Comparison of old vs. new radial velocity 
measurements, expressed in hundreds of km s$^{-1}$,
for 367 PNs in the W field. Since there is satisfactory agreement,
we have decided to average the two W measurements. See Figs. 2-4.
\label{fig1}}
\figcaption[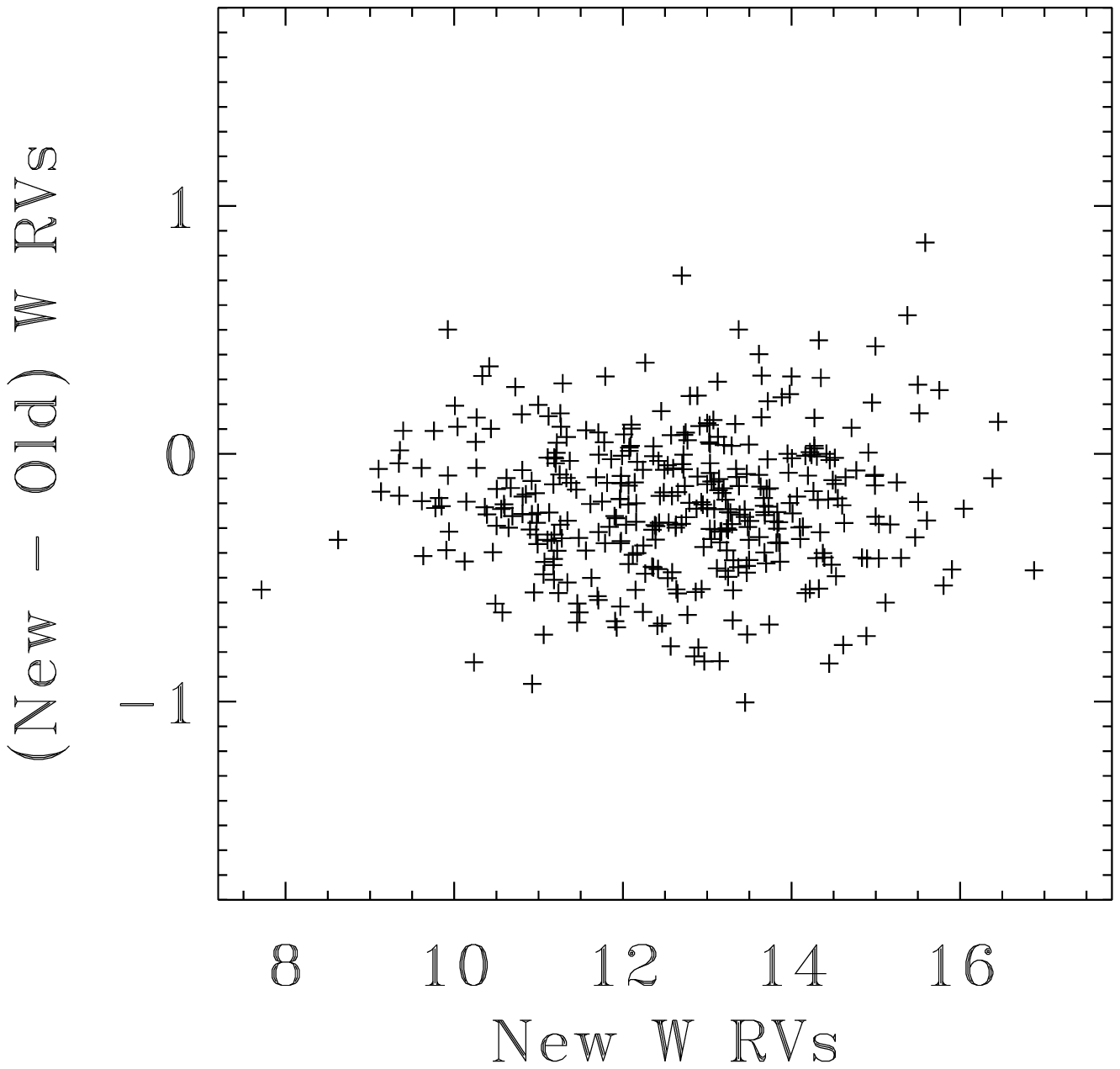]{Differences between new W and old W radial 
velocities as a function of the new W velocities.
The velocities are expressed in hundreds of km s$^{-1}$. 
The new W velocities are slightly lower, but
well within the errors of about 40 km s$^{-1}$.
\label{fig2}}
\figcaption[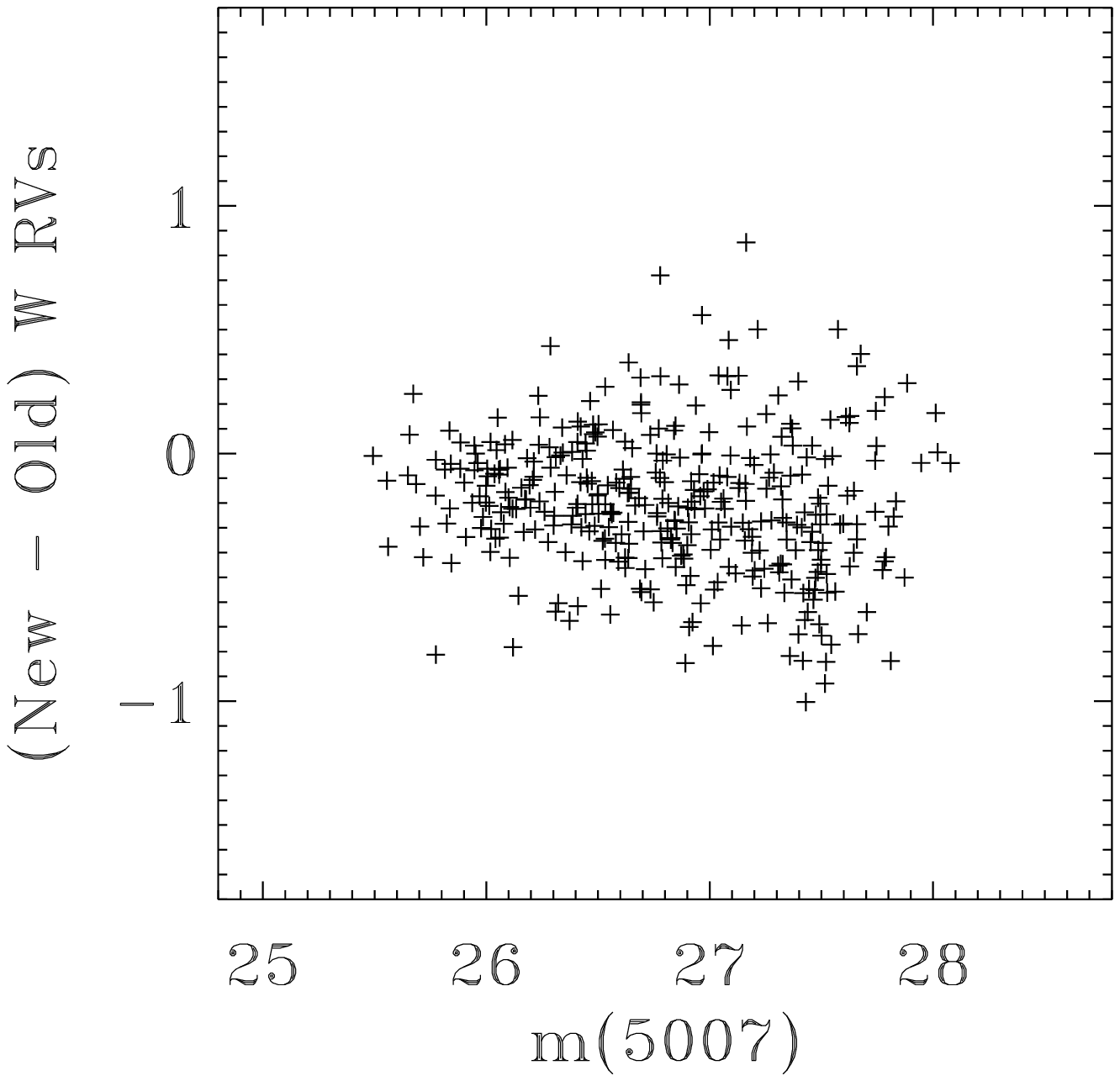]{Differences between new W and old W radial 
velocities as a function of $m$(5007). The velocities are expressed in 
hundreds of km s$^{-1}$. The new W velocities are slightly lower, but
well within the errors of about 40 km s$^{-1}$. The differences are 
larger for fainter PNs, as expected, since the quality of the radial 
velocities depends on the quality of the position measurements.
\label{fig3}}
\figcaption[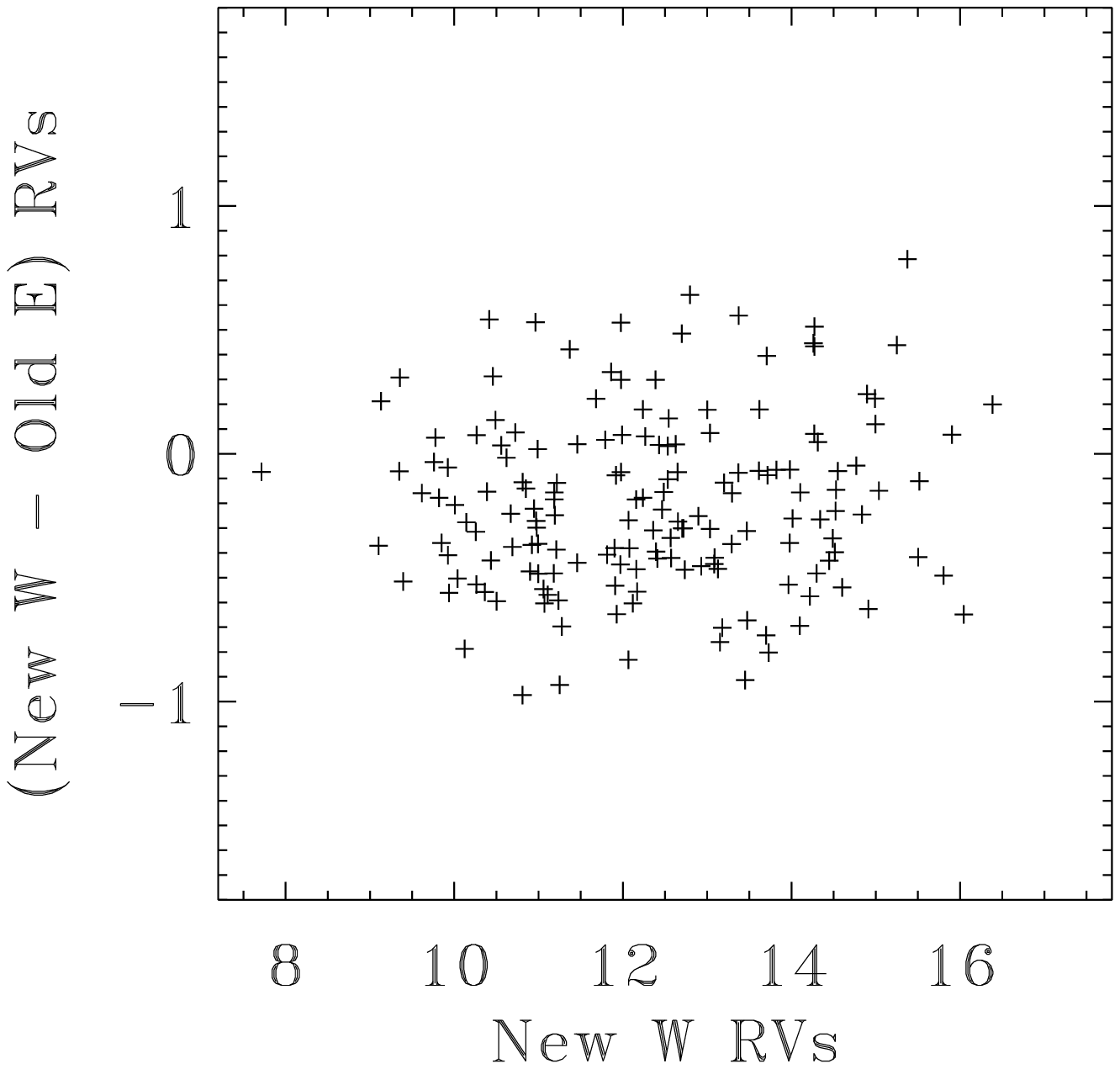]{Differences between new W and old E radial 
velocities
as a function of the new W velocities. There are 164 data points.
The velocities are expressed in hundreds of km s$^{-1}$. 
The new W velocities are slightly lower, but
well within the errors of about 40 km s$^{-1}$.
\label{fig4}}
\figcaption[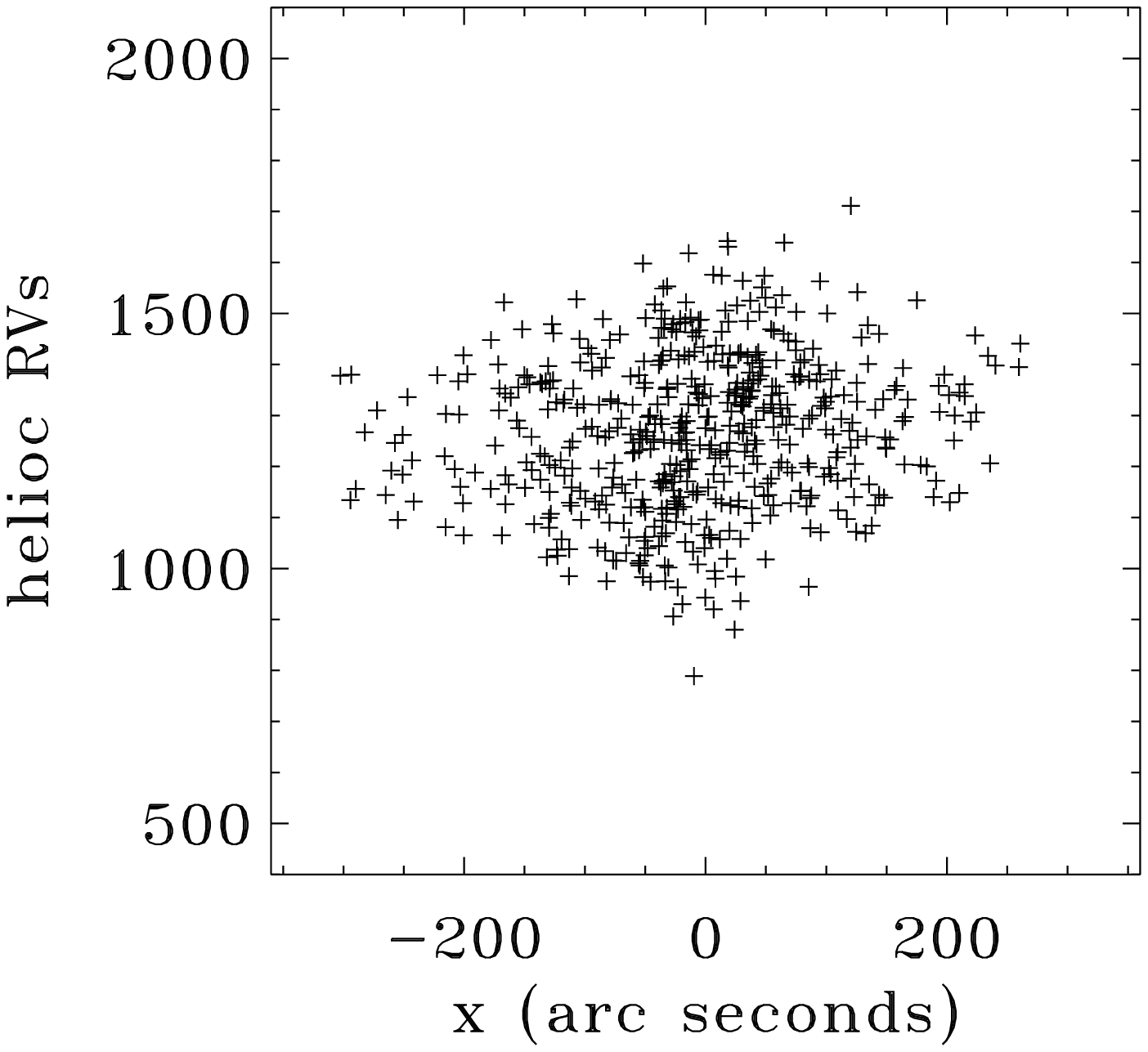]{Radial velocities of 531 PNs, in km s$^{-1}$, as a 
function of their $x$-coordinates in arcseconds relative to the center
of light of NGC 4697. The $x$-axis is oriented in the direction of the
major axis. This figure replaces Figure 20 in Paper 1.
\label{fig5}}

\clearpage

\begin{figure}
\figurenum{1}
\epsscale{1.0}
\plotone{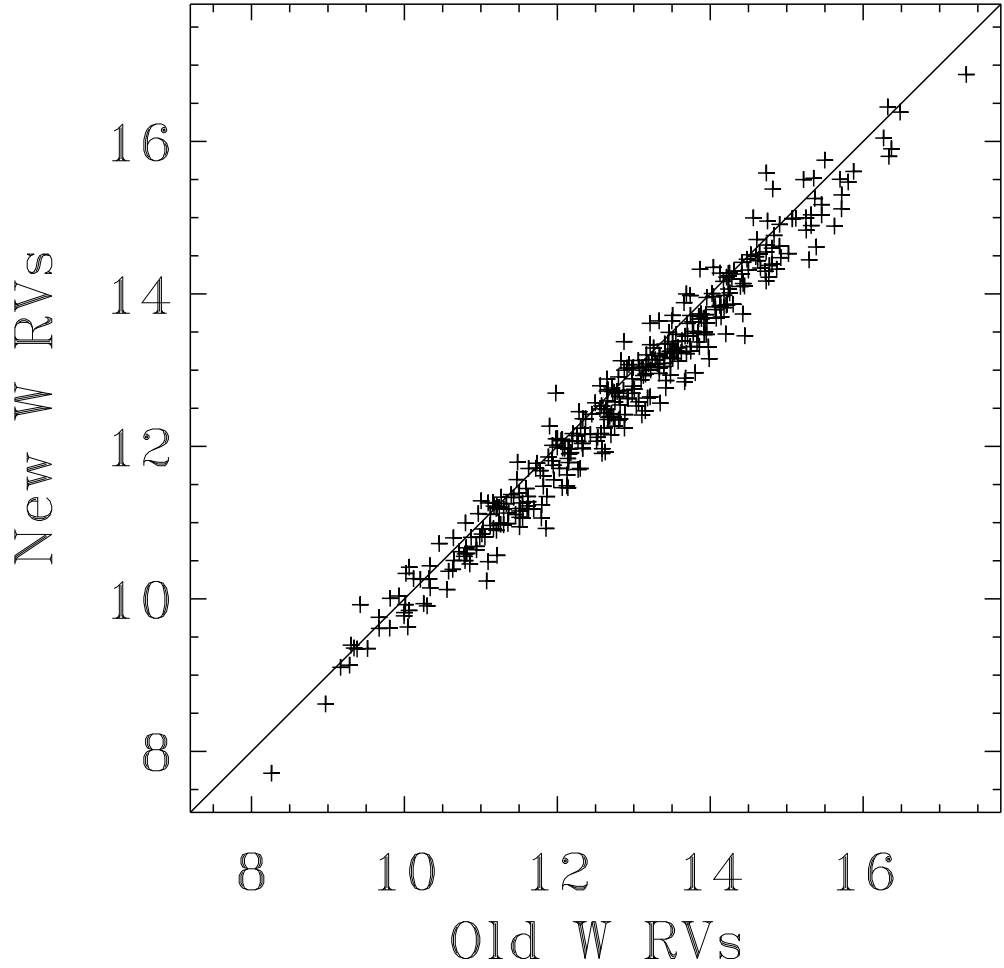}
\caption{Comparison of old vs. new radial velocity measurements,
expressed in hundreds of km s$^{-1}$, for
367 PNs in the W field. Since there is satisfactory agreement,
we have decided to average the two W measurements. See Figs. 2-4.}
\end{figure}

\begin{figure}
\figurenum{2}
\epsscale{1.0}
\plotone{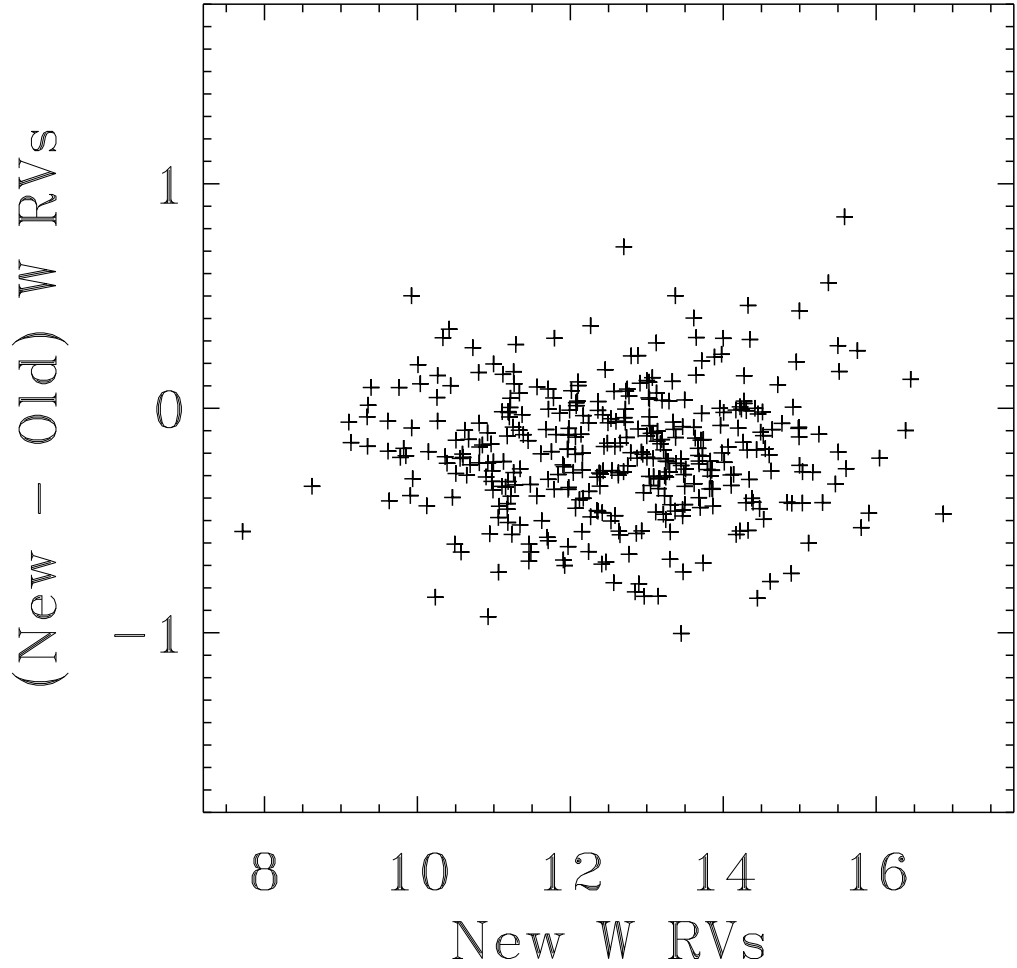}
\caption{Differences between new W and old W radial velocities
as a function of the new W velocities.
The velocities are expressed in hundreds of km s$^{-1}$. 
The new W velocities are slightly lower, but
well within the errors of about 40 km s$^{-1}$.}
\end{figure}

\begin{figure}
\figurenum{3}
\epsscale{1.0}
\plotone{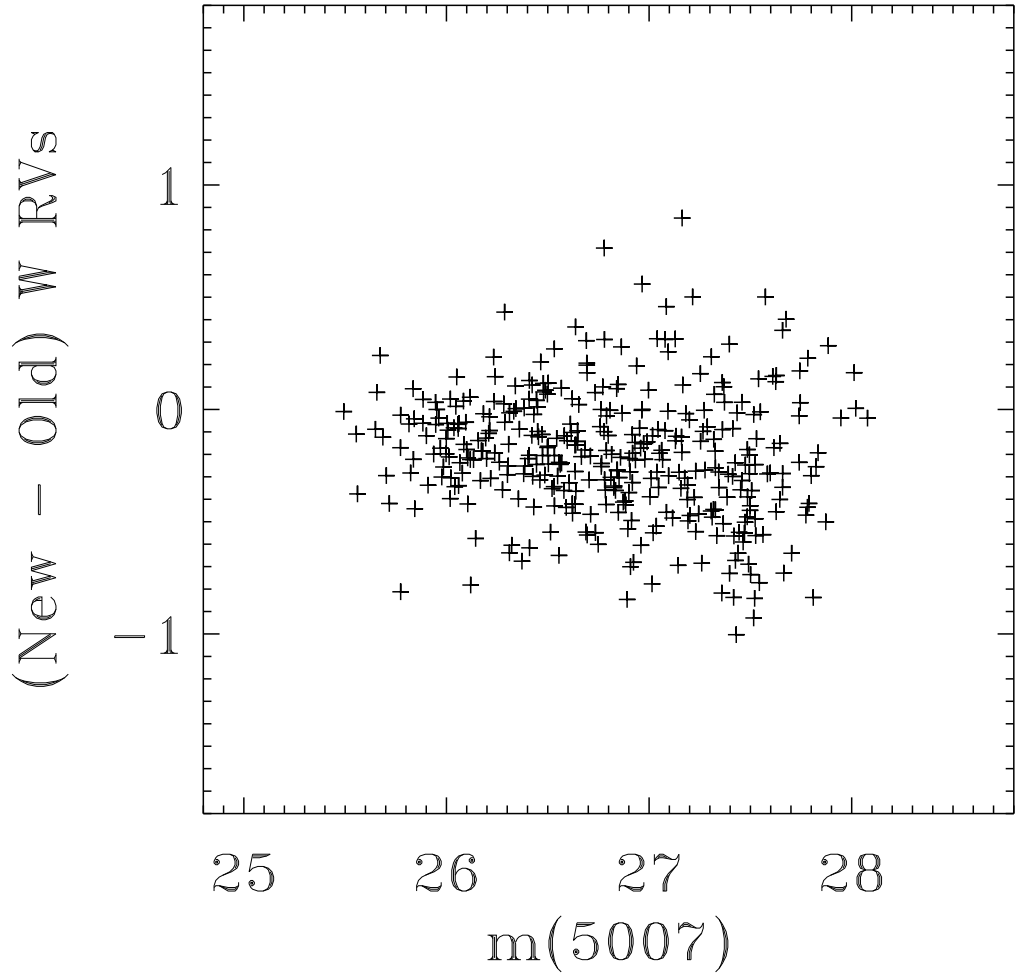}
\caption{Differences between new W and old W radial velocities
as a function of $m$(5007). The velocities are expressed in 
hundreds of km s$^{-1}$. The new W velocities are slightly lower, but
well within the errors of about 40 km s$^{-1}$. The differences are 
larger for fainter PNs, as expected, since the quality of the radial 
velocities depends on the quality of the position measurements.}
\end{figure}

\begin{figure}
\figurenum{4}
\epsscale{1.0}
\plotone{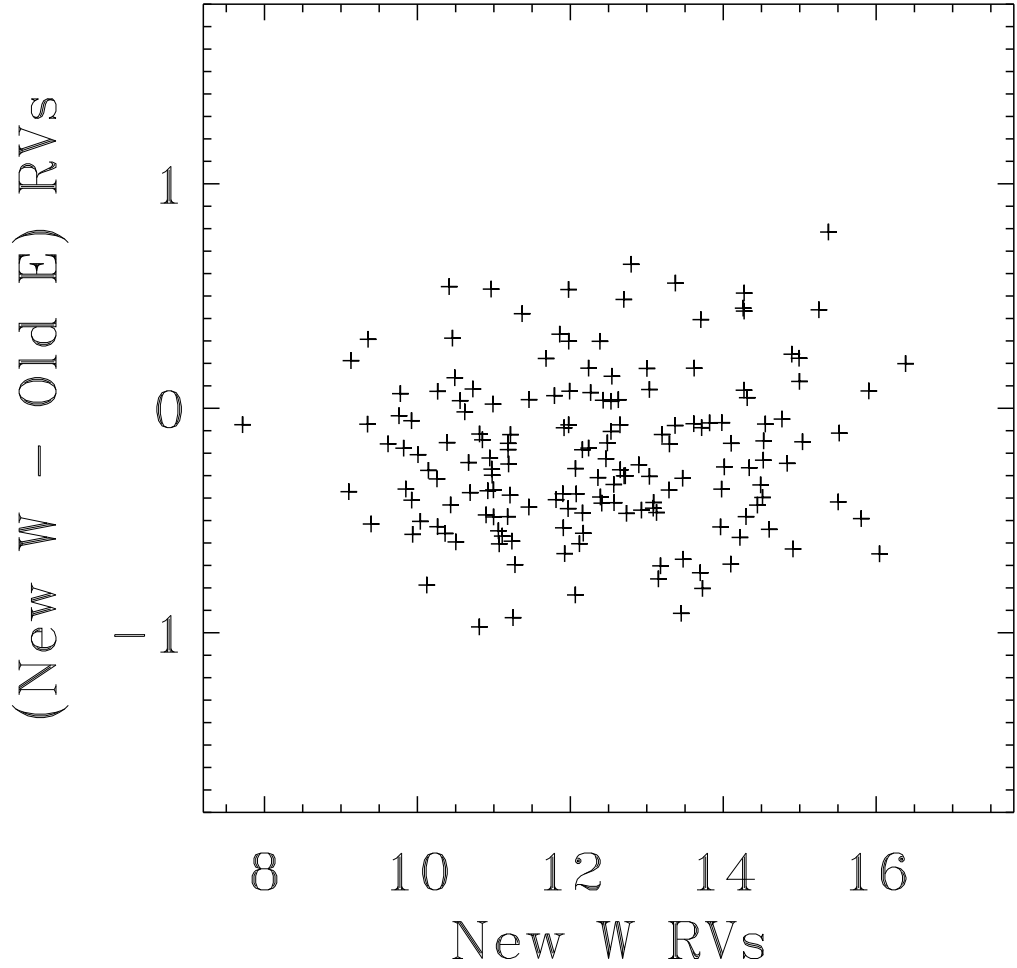}
\caption{Differences between new W and old E radial velocities
as a function of the new W velocities. There are 164 data points.
The velocities are expressed in hundreds of km s$^{-1}$. 
The new W velocities are slightly lower, but
well within the errors of about 40 km s$^{-1}$.}
\end{figure}

\begin{figure}
\figurenum{5}
\epsscale{1.0}
\plotone{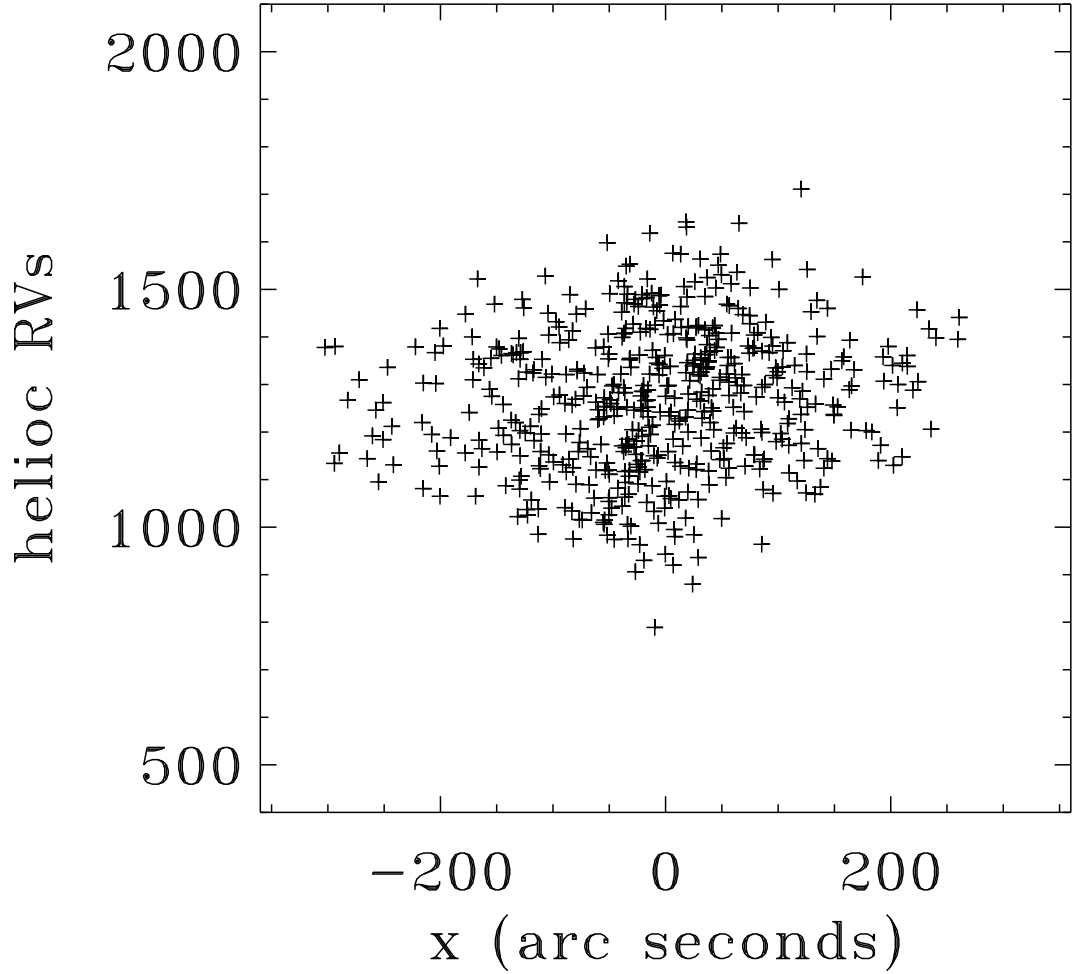}
\caption{Radial velocities of 531 PNs, in km s$^{-1}$, as a 
function of their $x$-coordinates in arcseconds relative to the center
of light of NGC 4697. The $x$-axis is oriented in the direction of the
major axis. This figure replaces Figure 20 in Paper 1.}
\end{figure}

\end{document}